\documentstyle[12pt]{article}

\def\title{
\bgroup%
\obeylines\everypar={\hskip\parfillskip}%
\large
\bf\vrule height1cm width 0pt\relax}
\def\endtitle{\vskip1sp\egroup}
\def\author#1{\hbox to\textwidth{\hss\vrule height.9cm width0pt\relax%
#1\hss}}
\def\contauthor#1{\hbox to\textwidth{\hss\vrule width0pt\relax%
#1\hss}}
\def\moreauthors#1{\hbox to\textwidth{\hss\vrule height.8cm width0pt\relax%
#1\hss}}
\def\instit{\bgroup\small\it\obeylines\everypar{\hskip\parfillskip}}
\def\endinstit{\vskip1sp\egroup}
\def \be {\begin{equation}}
\def \ee {\end{equation}}
\def \ba {\begin{eqnarray}}
\def \ea {\end{eqnarray}}

\begin{document}

\baselineskip 0.85 cm

\begin{titlepage}

\maketitle
\baselineskip 0.85 cm

\title{Critical behaviour of Random Walks}

\begin{center}

{\small\baselineskip=14pt I. Campos and A. Taranc\'on}

{\small\baselineskip=14pt {Departamento de F\'{\i}sica Te\'orica}
{Universidad de Zaragoza}
{Pedro Cerbuna 12,  50009 Zaragoza}
{Spain}}

\begin{abstract}

\baselineskip 0.85 cm

We have numerically studied the trapping problem in a two-dimensional lattice
 where particles are continuously generated. We have introduced interaction
 between particles and directionality of their movement. This model presents a
 critical behaviour with a rich phase structure similar to spin systems. We
 interpret a change in the asymptotic density of particles as a phase
 transition. For high directionality the change is abrupt, possibly of first
 order. For small directionality the phase transition is of higher order. We
 have  computed the phase diagram, the volume dependence of the critical point,
 and the relaxation time of the system in the large volume limit.

\end{abstract}

\end{center}

{\small\baselineskip=14pt {e-mail:}
 {isabel@sol.unizar.es}
 {tarancon@sol.unizar.es}}

\end{titlepage}

\newpage

\section{Introduction}
  One of the problems of a discrete-time lattice random
 walk is the trapping problem~\cite{CALDERON}. It concerns with the
temporal evolution of a system composed by N random walkers moving into a
random
 distribution of traps. This matter has been used to model
 diffusion processes~\cite{KUZOVCOV}~\cite{REVATHI}, adsorption of atoms in
 molecules, and magnetization decay in high $T_c$ superconductors where the
 particles are represented by vortices~\cite{SUPER}.

 We introduce an interaction, a so-called non-overlapping interaction, by
 forbidding occupation of a single site by more than one particle. We have also
 extended previous models by introducing directionality in the particle
 movement.
 Directionality is implemented by assigning  a final position represented by a
lattice site, uniformly distributed in the lattice, to every particle.
Particles
 will move to their final points according to a probability distribution
that is explained below. Due to
 directionality, the mean number of steps that a particle needs to reach its
 final point (at distance r) is not proportional to $r^2$, as in the
 free case~\cite{ITZYKSON}, but is proportional to
 $r^\gamma$ with $\gamma <2$.
 This holds for a system without interaction, where
 the  particles can be at the same lattice site. The interaction increases
when particle density increases.\\
 The conflict between directionality and density can be solved when the system
 relax. We distinguish two evolution possibilities: the system reaches
 saturation
which prevents the particles to move in any direction, or it reaches a
dynamical
equilibrium state. In the latter situation, each particle walks for some time,
 until it reaches its final point where it disappears. the density of particles
 is asymptotically stable.\\
 We have studied the transition between the dilute system and the saturated
 system. More precisely if this change is abrupt or continuous, i. e. if there
 is a phase transition between them, and  what could be the order of this
 transition. A rich phase diagram appears where we distinguish two regions, one
 of them with a first order transition, and another one with a higher order,
 possibly second order.
 We have calculated the behaviour of the system when
$V\longrightarrow\infty$.\\
The time in which the asymptotic state is reached, the relaxation
time $\tau_r$, strongly depends on the directionality and density. The
behaviour
 of $\tau_r$ is similar
to the behaviour of the correlation length close to a phase transition in a
spin model and to the magnetization relaxation of a High  $T_c$
 superconductor~\cite{SUPER}.

\section{Model}

 Consider a two-dimensional discrete-time lattice where
 lattice every point is neighboured by four others. We impose periodic
 conditions
 in both directions. Particles sit on lattice sites $n$, that are characterized
 by $(n_0,n_1)$.\\
 The model can be described with the two parameters $\beta$ and $p$.\\
 For every time step, we consider all lattice sites. If the considered site
 is free, we create a particle with creation probability $p$. If the particle
is
 created, a final point for this particle is generated uniformly in the
lattice.
 If the considered site is occupied, the particle is moved in the direction
 $\mu$,
 ($\mu$ takes the values +0,-0,+1,-1) according to the probability distribution
\be
P(\pm \mu) = N e^{\pm (sig (n^f_\mu - n_\mu) \beta )}
\label{prob}
\ee
where $n^f_\mu$ is the coordinate $\mu$ of the final point of the particle,
$n_\mu$ is the coordinate  $\mu$ of the particle, and sig(x) is the $signum$
function. $N$ is chosen to normalize the probability:\\
\be
\sum_{\pm\mu} P(\mu) = 1
\ee
When the chosen site is occupied, the particle does not moves and it remains at
 the same lattice site until the next time step.
 If the site is free and it is precisely the final point, the particle
 disappears.\\
 A particle can at most move once in one time step.\\
  To implement this, we assign to every site a "spin" taking the value 1, if
 the site is occupied, or 0 if it is free.\\
 We describe the evolution of the system trough the mean
 number of particles. By analogy with spin systems,  this number is called the
 magnetization of the system
\be
M=(1/V){\sum_{{\bf i=1}}}^V \sigma(i)
\label{mag}
\ee
 where $\sigma(i)$ equals 1 if site is occupied, and 0 if site is
 free. The magnetization is to the density of particles in the system.\\
 The choice of the parameters $p$ and $\beta$ is due to
 the fact that these parameters present limits easy to recognize.
$\beta$ can be regarded as the inverse of the temperature of the system.
 Movements of the particle are more random when the temperature is increased.
We
 will return to this point below.\\
 It is interesting, and very intuitive, to recognize what happens for limiting
 values of the parameters. When $\beta\longrightarrow 0$, the probability takes
 the
 simple form $P(\mu)=1/4$ which corresponds to a random choice of the next
site.
 In the absence of other particles this is like a brownian
 movement of particles with a "high temperature".When
 $\beta\longrightarrow\infty$, the particle can only move in
 the direction  $\mu$ which brings it nearer to its final point. In this case
 there
 are no alterations from the minimum-length path that joins the particle
 with its final point. Thus the system is at "zero temperature".\\
 In the limit $p\longrightarrow 0$, the system has a very low particle density,
 and presumably little interaction between the particles. The particles move
 from their origin to its final point with little fluctuations depending on
 $\beta$.
When $\beta$ aproaches 0, we see a typical random walk  of particles
that are generated with probability $p$. The number of surviving particles
after
 n-steps is related with the expected number of distinct sites visited by an n
 step random walk~\cite{WEISS} -~\cite{PEACOCK}.
 The particles arrive at their final site, at distance
 $r$, after $r^2$ steps. When increasing $p$, the system becomes more dense.
Due to
 the interaction between particles, the number of steps that a particle needs
to
 reach its final point increases. Ultimately there is a situation in which the
 system saturates, when the density becomes high enough to prevent any
movement.
 No particle
 can move through the lattice and the temporal evolution of the
system will not change significatively. Now, we increase the value of $\beta$,
so that, directionality increases and the particle arrives faster at its final
 point. Thus the value of $p$ that produces saturation would be larger.\\
 We distinguish other regions in our model. When $\beta$ is large enough,
the directionality is so strong that particles always choose the minimum path
to
reach their final point. This phenomenon prevents the overlap between the path
of particles with opposite final points. The system saturates for smaller $p$
 values. \\
 About the transition, we would like to point that for small values of $\beta$,
the system can
 reach large values of magnetization without system saturation
 because the interaction is of little importance, so that a particle can reach
 its final site
 through various ways. Before saturation, the density is large, and the
 transition between the two phases is not abrupt.\\
  On the other hand, for large $\beta$ (but not large enough to produce the
 no-overlap between paths explained below) every particle moves to its final
 point quickly. For small values of $p$, the density is small. When increasing
 $p$, particles increasingly interact and the system creates obstacles that
 propagate quickly. The system saturates more abruptly than for small
$\beta$.\\

\section{Numerical Simulation}
 We have run Monte Carlo simulations in order to study the full parameter space
 of the model, the possible phase transition, and the behaviour in the
 $V\longrightarrow\infty$ limit, and the relaxation time.\\
  The two-dimensional square lattice has a length $L$, with periodic boundary
 conditions .\\
 Simulation has been performed by using a custom machine with 64 transputers
T805~\cite{RTN}.\\
 We have mapped out the $(\beta, p)$ parameter space globally for some $L$
 values to find strong changes in the temporal evolution of the system.
 Time ($t$) is expresed in the number of iterations of the
 Monte Carlo simulation, where the iteration is a tentative sequential update
of
 all lattice sites.
 The updating procedure for a single site has the following steps:\\
\begin{itemize}
\item {A.} If the site is free, generation of a random number  between 0 and 1
 occurs. If the randomly number generated is less than $p$, a particle appears
 at this site. Otherwise, no particle is created. In order to implement the
 "occupation" of a site, we assign a "spin" of value 1 to the site if the site
 is occupied, and 0 if the site is not occupied. In this way, we can obtain the
 number of particles in the system by summing all "spins".
\item{B.} If the site considered is occupied by a particle, this particle
 attemps to move in one of the four sens according to the following: Since we
 know the
 probability for all directions from (\ref{prob}), we
 divide the interval [0,1] in four parts proportional to the probability
 assigned to every spatial direction. Next, a random number between 0 and 1 is
 generated. This number is contained in one of the four intervals in which the
 interval [0,1] is divided. The particle will move in a direction corresponding
 to that interval. With this algorithm, we reproduce the probability
 distribution weighted by $\beta$.
\begin{itemize}
\item{B1.} If the chosen site is not occupied, the particle moves to
 it, and when that site is the final point of the particle, it disappears at
 this same step.
\item{B2.} If the site is occupied, the particle remains at its original point,
 and the next site in the lattice is considered.\\
 The interaction between particles is implemented
 by preventing a particle to move into an occupied site.\\
\end{itemize}
\item{C.} Particles move at most once per update.
\end{itemize}
When the final point is at a distance larger than $L/2$, considering distance
as
 $x_{max}-x_{min}$, because of the periodic boundary conditions the most
 favourable direction is such that the particle takes the shortest path.\\
The asymptotic value of $M$,
$$
M_\infty= \lim_{t\to\infty} M(t)
\label{minf}
$$
is the parameter which labels the phases of the model.
The relevant quantity in order to study the situation of the system is the
 magnetization as defined in (\ref{mag}). $M$ depends on $t$ (t is the number
of
 Monte Carlo iterations) and if the $M$ goes approaches one for
 $t\longrightarrow \infty$, the
system is saturated, the number of particles that arrive to their final point
 does not compensate the rate of creation. If the $M < 1$
 system has reached a dynamical equilibrium between creation and annihilation
of
 particles.\\
$M_\infty$ is not a true order parameter from a statistical point of view,
 because it is always positive over all the phase diagram. However, as we will
 see in the next sections, this quantity presents a clear change of behaviour
as
 a function of $p$ at each value of $\beta$, and it is therefore a good
 parameter in order to establish where the transition takes place. We call the
 value where this change on $M_\infty$ appears, $p_c$. In fact $p_c$ a is
 function of $\beta$ and $L$, the lattice size.

\section{ Results and Phase Diagram}

\subsection{General results about the phase diagrams}

At each value of $L$, $\beta$ and $p$ we run a Monte Carlo simulation. The
 starting configuration is obtained by creating a particle at each site with
probability $p$. We start therefore with an initial magnetization of order
$pV$.

We have studied several lattice sizes: $L=20,40,60,80 and 100$. We have run a
different number of Monte Carlo iterations depending on the convergence of
$M(t)$. At each value of the parameters, we have performed typically
$1-2 \times 10^5$ MC iterations. We have also made the simulation starting from
differents configurations,  and we have allowed the system to evolve until
$5 \times 10^5$ iterations, in order to accurately compute the value of
 $M_\infty$. We have computed the error in $M_\infty$ and $p_c$ by calculating
the dispersion between the results obtained starting from differents
configurations.\\

At each MC iteration we measure the number of particles (\ref{mag}), obtaining
 $M(t)$, a function of the MC time $t$ (see Fig. 1). Our order parameter, in
 order to determine the phase we are in, is the asymptotic value of this
 parameter $M_\infty$. As can be seen in this figure, the time in order to find
 this asymptotic value depends strongly on $p$. At each $\beta$ value in the
 small $p$ ($p<<p_c$) region ("dilute system"), this behaviour is reached
 after a small time an also in the large $p$ ($p>>p_c$) region. In the
 intermediate $p$ ($p\approx p_c$) region (where the system undergoes very
 strong changes), this time becomes enormous (see 4.3).\\
 We have carried out a careful study in order to control that in every case the
 asymptotic behaviour is reached and further time evolution does not change
 $M(t)$. Firstly from different starting configurations, to see if the final
 state depends on the initial one. We have found no evidence of this
 dependence. In fig 2, we show $M(t)$ for several starting configurations.
 Secondly after $M(t)$ has reached an apparently asymptotic behaviour, because
 the time evolution of $M(t)$ seems to be stable, we have completed up to 4
 times more iterations ($5 \times 10^5$ iterations) $M(t)$ continues stable.\\
In Fig. 3 we present the evolution of $M_\infty$ as a function of $p$ for
 two values of $\beta$ and $L$ fixed.

 As can be observed in this figure, the system presents two phases. There is
 phase in which the system is saturate, and another one in which asymptotic
 magnetization is less than 1.\\
First, for small values of $p$, there is no saturation. In this case, the
 density is small and the system never collapses. We go through the parameter
 space by increasing $p$. For $p = p_c$, the system changes from a
 value of $M_\infty$  less than 1, to be saturate .\\
However the behavior of this quantity is different for large and small $\beta$.
 The system makes this change in a smooth way for small values of $\beta$, and
the change is less smooth when $\beta$ increases. For $\beta$ close to
$1$, the change is abrupt, the density changes from a small value to $1$, by
making a jump.\\
For large $\beta$ values the strong change at $p_c$, tells us that $M_\infty$
is
 discontinuous and we can speak of a behaviour similar to a first order
 transition as in $Z_2$ Spin Gauge Systems at high dimensions. As we will see
we
 can find
 also a quantity with a similar behaviour to the correlation length in the spin
 systems, making this similitud deeper.

For smaller values of $\beta$, also both phases are present, but now the
 change is smoother, and we can see that there is no discontinuity in
 $M_\infty$.
However for large $p$, $M_\infty=$Constant$=1$, and for small $p$, $M_\infty\to
 0$, therefore a discontinuity must occur at some
$(\partial^n M/\partial^n p)$ at some intermediate value of $p$; this is
similar
 to higher order phase transitions, where discontinuities at the $n$ derivative
 of operators is a signal of a $(n+1)$ order transition.

 In the phase diagram on Fig. 4 we plot the values obtained for $p_c$ for
 wide range of $\beta$ values. We can see how $p_c$ increases until a maximum.
 After this maximum $p_c$ decreases strongly and for large $\beta$ , $p_c$
decays slowly (Fig. 4) \\
 We try to explain this behaviour: At first, when $\beta$ increases, particles
arrive more quickly to their final point. Because of this the system supports
the $p's$ larger without saturation. This fact occurs for the first region of
$\beta$.
 When $\beta$ has increased enough, particles can't go away their shortest path
 to reach the final site. A particle cannot avoid other particle by going in
the
 opposite sense. This produces a collapse that propagates quickly to the entire
 lattice. The value of $p_c$ decreases for large $\beta$ due to this
phenomenon.

\subsection{Volume Dependence}

 We have performed the simulation for various lattice volumes. As can be
 observed in the figures, for fixed value of $\beta$, the value of $p_c$
 decreases when the lattice size increases. This behaviour is expected because
 particles must walk larger distances when the lattice size increases.
Therefore
if the path is larger, a particle will find more particles in its way
 ("obstacles") to reach the final point. These two facts combine and
 $p_c$ decrease. We can summarize this by saying that saturation is produced
 earlier when lattice size increases.\\
  The volume dependence of $p_c$ for a given value can be studied with accuracy
for those values of $\beta$, in which the transition is abrupt, because $p_c$
is
 determined more precisely. We have studied this dependence for
 $\beta=1.4$ and $\beta=2.0$. Results can be observed in the figure 5. We find
 that in both cases the dependence can be ,written in the form:
\be
p_c= A L^\alpha
\label{alpha}
\ee
where $\alpha=2.33 \pm 0.01$ for $\beta=1.4$, and $\alpha=2.36 \pm 0.01$ for
 $\beta=2.0$.\\
We have checked that this power law is true for all values of $\beta > 1.0$

The fact that $\alpha>2$ in (\ref{alpha}) is important in order to establish
the
 $L\to\infty$ limit of the model in the particular case of constant number of
 particles.
That is to say: consider the special case where the same number of particles is
created independently of $L$. To do that, we must fix $p$ proportional to
 $L^{-2}$.
At small $t$ values, where system is not saturate even in the $p>p_c$ region,
 the number of particles created at each time is of the order of
$$
pL^2(1-M(t))\propto pL^2
$$
valid if $M(t)\ne 1$. In the case of constant rate of particle creation
 independently of $L$, the situation in the large $L$ region is very different
 depending on $\alpha$.

If $\alpha>2$, the system has only one saturate region in that limit, and
if $\alpha<2$ only dilute region is present. The third case is when $\alpha=2$
in which case both phases exist.

In general, for any finite $L$ two different phases are present in all cases.

For small $\beta$ values, there is great difficulty to find the critical point
accurately. This is due to the fact that saturation is reached at high values
 of $M$. So, we have not found a clear power law in this region.

\subsection{Relaxation Time}

 As we have pointed out, the system needs a time in order to reach the
 asymptotic value of $M$. We have studied the evolution of this time with $p$,
as a function of $L$ and $\beta$. To be more definite we have measured, for
 every value of the parameter $p$, the time the system needs to reach a value
of
 $M$ equal to $M_\infty/e$, called relaxation time $\tau_r$. \\
  We expect smalls values of $\tau_r$ for values of $p$ far from $p_c$.
  For $p$ in the neighborhood of $p_c$, $\tau_r$ increases, and will
 reach its maximum value for $p=p_c$.\\
 We have supposed that the temporal evolution follows an exponential law:\\
\be
M_\infty-M(t) = Be^{(-1/{\xi)}.t}
\label{cor}
\ee
 With this definition, the relaxation time, $\tau_r$, is given by $\xi$.

We found a good agreement of our data with this formula.
In figure 6 the result for a lattice size $L=40$, with $\beta=1.0$ can be
observed.
 We have plotted in the little window the temporal evolution for a value of
$p > p_c$. We have obtained the value of $\xi$ by plotting the $\ln(M-m(t))$,
 fitting to an straight line whose slope is $1/\xi$:
\be
\ln(M_\infty-M(t)) = A + (1/{\xi}).t
\ee
 The fit to a straight line is good for $p > p_c$, because the behaviour of
 M is better defined. M does not fluctuate when the system reaches
 saturation. For small values of $p$, system is less "determinate", M
fluctuates
 around its mean value, and the fit is less accurate.
 Finally, we have represented the evolution of $\tau_r$ for various values of
 the parameter $p$, finding the expected behaviour (see figure 7).

The correlation $\xi$ has a behaviour similar to that of the correlation length
in a spin system around a phase transition. For large values of $\beta$, for
$p$
over $p_c$ but close to it, correlation is little. For $p$ less than $p_c$, but
close to it, correlation is also little. This is due to the fact that in both
cases the system reaches the asymptotic behaviour in a small time. In fact, the
transition is discontinuous and $M_\infty$ is lower than 1 for $p<p_c$ and
$M_\infty \approx 1$ for $p>p_c$ (fig 3). At $p_c$, $M_\infty$ depending
on the starting configuration, the system evolves to the saturation or to the
dilute regime, in a similar way to a spin system at the critical temperature on
a first order phase transition.

For $\beta$ small, calculation of $p_c$ has a very high error, and measure
accurately $\tau_r$ has no been possible. With the partial obtained results
a possible scenario could be the following:

For small values of $\beta$, near $p_c$, $M_\infty$ is continuous, approaching
1
 for $p \longrightarrow p_c$ (fig 3). For $p<p_c$, the system evolves in a
finite time to $M_\infty$, and $\tau_r$ is finite. Also for $p>p_c$. At
$p =p_c$ the system needs and infinite time to reach $M_\infty$, and then
$\tau_r$ diverges. This behaviour is similar to a second order phase
transition.

{\bf Acknowledgements}
 We thank CICyT AEN93-0604-C03-01 partial financial support and the
RTN Group for the use of the RTN machine where part of this
computation has been performed.\\
 The authors thank Silvia Mili\'an and Peter Booi their carefully reading
and  their useful suggestions.

\newpage
{\Large Figure Captions}
\bigskip

\begin{enumerate}
\item Temporal evolution of magnetization for $p<p_c$ and
      $p>p_c$ performed for a $L=40$ lattice with $\beta=1.0$.
      $p_c$ is $0.00035$
\item Temporal evolution of $M$ starting from three different configurations
for
      $L=40$, $\beta= 1.0$ and $p=0.00025$.
\item Asymptotic magnetization for $p\approx p_c$,in a $L=40$ lattice with
      $\beta=0$ and for $\beta=1.4$.
      The discontinuity is abrupt for $\beta=1.4$, and is smooth for
      $\beta=0.2$. The value of $M_\infty$ is calculated after
      $3-4 \times 10^5$ MC iterations.
\item Phase diagram for L=20,40,60 and 100. We have used logarithmical scale
      to represent $p_c$
\item Volume dependence of
      $p_c$ for $\beta=1.4$ and for $\beta=2$. Points have been fitted to a
       straight line. The fit is good in general for $\beta > 1.0$.
\item $M_\infty-M(t)$ for $L=40$, $\beta=1.0$,
      $p=0.00045$. Points have been fitted to a straight line whose slope is
      $\xi$, which acts as a "correlation length".
\item  Evolution of $\tau_r$ with $p$, for $\beta=1.0$ in a $L=40$ lattice.
       We have plotted $\tau_r$ as a function of $(p-p_c)/p_c$, called $p_r$.
       $\tau_r$ is maximal for $p_r = 0$, id est for $p=p_c$.

\end{enumerate}


\newpage

\begin{thebibliography}{99}


\bibitem{CALDERON}
C.P. Calderon and T.A. Kwembe,
{\sl Mathematical Biosciences} {\bf 102, pp 183-190} (1990)
\bibitem{KUZOVCOV}
V. Kuzovcov and E. Kotomin,
{\sl Rep. Prog. Phys.} {\bf 51 pp 1479 } (1988)
\bibitem{REVATHI}
S. Revathi and V. Balakrishnan
{\sl J. of Physics A} {\bf 26 p 257} (1993)
\bibitem{SUPER}
O.F. Schilling,
{\sl Phys. Rev. B} {\bf 44 6 pp 2784-2788} (1991)
\bibitem{ITZYKSON}
C. Itzykson, J.M. Drouffe
{\sl "Theorie statistique des champs". Chapter I } {\sl Intereditions du CNRS}
(1989)
\bibitem{WEISS}
G.H. Weiss, I. Dayan, S. Havlin, J.E. Kiefer, H. Larralde, H.E. Stanley and P.
Trunfio
{\sl Physica A} {\bf 191 pp 479-490} (1992)
\bibitem{DVORETZKY}
A. Dvoretzky and P. Erdos
{\sl Proc. Second Berkeley Symposium. pp 33}  {\sl Univ. of California Press}
(1951)
\bibitem{SHEU}
W.S. Sheu and K. Linderberg
{\sl Physic Letters A} {\bf 147 pp 437-441} (1990)
\bibitem{GWEISS}
S. Havlin, G.H. Weiss and R. Kopelman,
{\sl Physic Rev. A} {\bf 39 p 466} (1989)
\bibitem{PEACOCK}
E. Peacock-Lopez and J. Keizer
{\sl J. Chem.Phys.} {\bf 88 p 1997} (1988)
\bibitem{RTN}
RTN Collaboration
{\sl Proc. CHEP 1992 Conference} {\bf CERN 92-07)}
\end{thebibliography}
\end{document}